\documentclass[twocolumn,showpacs,preprintnumbers,amsmath,amssymb,superscriptaddress]{revtex4}
\usepackage{epsfig, amsmath, times, bm, dcolumn, mathptmx, graphicx}

\begin{document}
%\preprint{to be submitted to Physical Review Letters}
\title{Observation of single defect relaxation in a freely suspended nano resonator}
\author{Florian W. Beil}
\address{Center for
NanoScience and Sektion Physik, Ludwigs-Maximilians-Universit\"at
M\"unchen, Geschwister-Scholl-Platz 1, 80539 M\"unchen, Germany.
}%
\author{Robert H. Blick}
\address{Electrical and Computer Engineering, University of
Wisconsin-Madison, 1415 Engineering Drive, Madison WI 53706, USA.
}%
\author{Achim Wixforth}
\address{Lehrstuhl f\"ur Experimentalphysik I, Universit\"at
Augsburg, Universit\"atsstra{\ss}e 1, 86135 Augsburg, Germany.
}%
\author{Werner Wegscheider}
\address{Institut f\"ur Angewandte und Experimentelle Physik,
Universit\"at Regensburg, 93040 Regensburg, Germany.
}%
\author{Dieter Schuh}
\address{ Walter Schottky Institut, Am Coloumbwall 3, 85748
Garching, Germany.
}%
\author{Max Bichler}
\address{ Walter Schottky Institut, Am Coloumbwall 3, 85748
Garching, Germany.
}%
\date{\today}% It is always \today, today,
\begin{abstract}
Relaxation of single defects in a nanometer sized resonator is
observed by coupling surface acoustic waves to a freely suspended
beam. The surface waves act on the resonator as driving forces
being able to modify the internal friction in the beam. In analogy
to classical experiments on internal friction in macroscopic
samples, we perform frequency, amplitude, and temperature
dependent experiments on the nano resonator and find a scenario
which is consistent with the observation of single defect
relaxation.
\end{abstract}
\pacs{85.85.+j, 43.35.+d, 61.72.Hh}% PACS, the Physics and Astronomy
\maketitle
%%%%%%%%%%%%%%%%%%%%%
The motion of dislocations in elastic solids under periodic load
is known to contribute to losses occurring in internal friction
experiments~\cite{Fantozzi:1}. Therefore, internal friction
experiments offer a method to explore the dynamics of
dislocations. This is in particular interesting if the elastic
body under investigation is very small, not much larger than the
typical dimension of the dislocation itself. The profound
knowledge of the dominating energy dissipation mechanisms in
nanometer sized resonators is in particular important in view of
the visions using such systems as narrow band high frequency
filters~\cite{Nguyen:1} or for future sensor
applications~\cite{Cleland:1}. To date, mechanical resonance
frequencies of nanometer sized resonators up to 1~GHz have been
reported~\cite{Roukes:1}. Unfortunately the dramatic reduction in
resonator size over the last few years turned out to be
accompanied by two undesirable facts. While nano crystalline
materials seem to supply superior mechanical properties, like
increased yield stresses in fine grained metals described by the
Hall-Petch relationship~\cite{Wang:1}, the quality factors of nano
crystals under resonant excitation considerably decrease with
size. For these reasons, it would be highly desirable to study the
internal losses of a nano resonator under conditions where defect
relaxation becomes observable. %So far, however, the lack of
%appropriate mechanical excitation mechanisms for the high resonant
%frequencies and small dimensions of a nano resonator hampers the
%use of classical methods to investigate the damping mechanisms
%active on the nanometer scale.
\begin{figure}
\includegraphics[width=8.5cm]{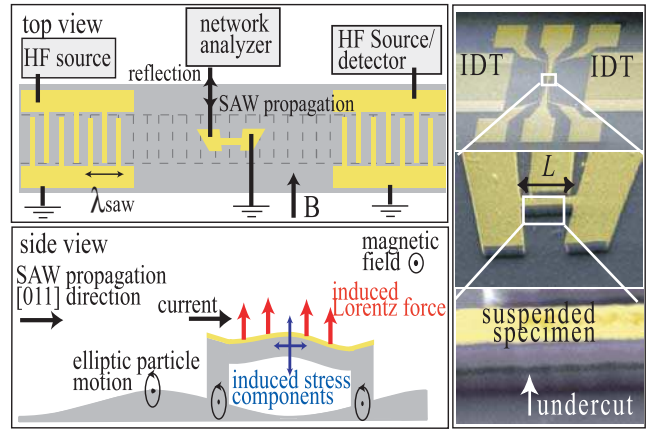}% Here is how to import EPS art
\caption{left: Top and side view of the experimental setup: The
nano crystal under investigation is placed in the sound path of
two interdigitated transducers generating the surface acoustic
waves. The wavelength $\lambda_{\rm saw}$ is given by the
lithographically defined finger spacing of the IDT. right:
Electron micrograph of a typical sample showing the freely
suspended nano crystal of length $L$ in the center of the acoustic
delay line. The SAW center frequency is 305~MHz, corresponding to
a wavelength of $\lambda_{\rm saw}$ = 9~$\mu$m. The beam length
$L$ = 3.4~$\mu$m is matched to $L = \lambda_{\rm saw}/2$, taking
into account additional underetching of the suspensions. The beam
width and height are w=300~nm, and h=200~nm, respectively. }
\label{figure:1}
\end{figure}

\begin{figure}
\includegraphics[width=8.5cm]{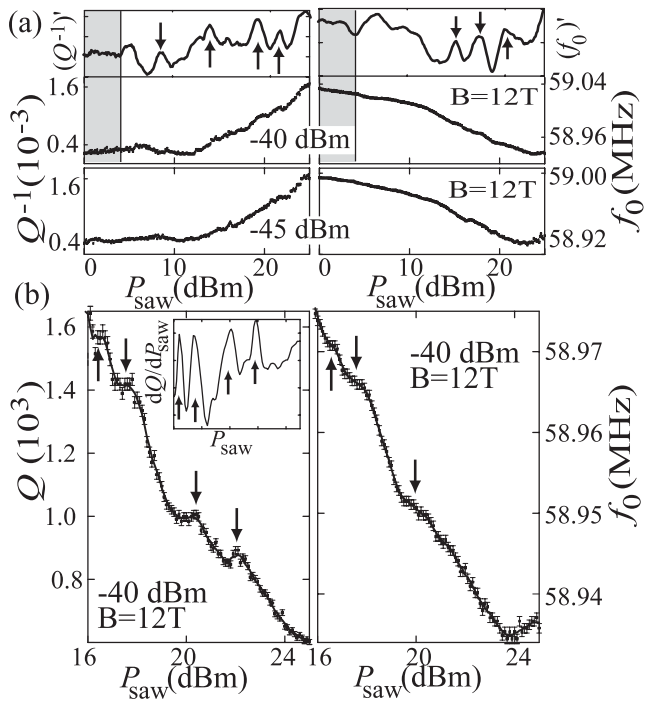}% Here is how to import EPS art
\caption{(a) Tuning of the eigenfrequency and damping of the beam
resonator with increasing SAW power for different probe powers at
the nano crystal. The upper curves in the -40~dBm traces show the
smoothed derivatives of the corresponding data. For low SAW power
(grey area) no tuning is observed, whereas above a certain
threshold power damping increases. (b) Magnified view of the
dependence of quality factor and eigenfrequency on SAW power. The
applied probe power to the beam was $P_{\rm res}$=-40~dBm at
B=12~T. The error in the determination of $Q$ is on the order of
2\%, as depicted by the error bars. The inset in the $Q$ trace
shows a smoothed derivative of the data.} \label{figure:3}
\end{figure}

\begin{figure}
\includegraphics[width=8.5cm]{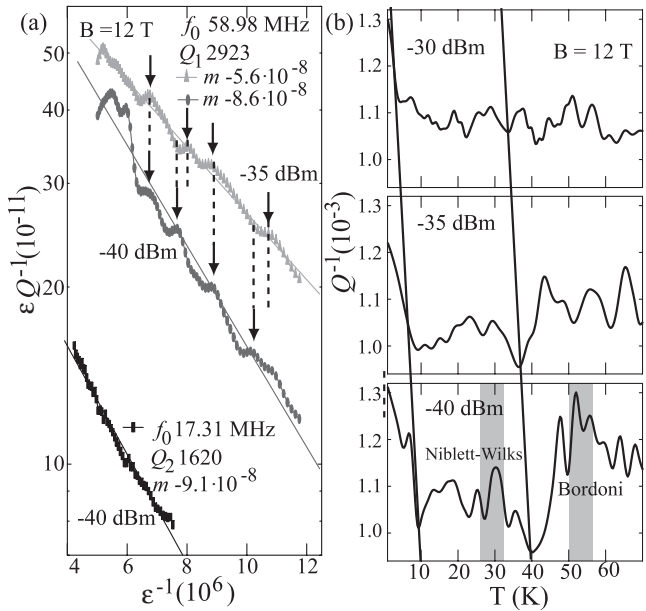}% Here is how to import EPS art
\caption{(a) Granato-L\"ucke plots for the $Q$ tuning as shown in
Fig.~\ref{figure:3} (upper two curves). With increasing power at
the beam the decreasing slope $m$ indicates an increased average
loop length due to unpinning by increased forces acting on the
nano crystal. The curves exhibit an overall linear dependence
modulated by accentuated steps. The steps disappear for another
sample, with higher dislocation density and increased amount of
pinning points. (b) Dependence of the internal friction on
temperature for different probe powers without acoustic loading.
The traces exhibit a peak at T$\sim$30~K, followed by an increase
in friction. The shaded regions in the -40~dBm measurement mark
the expected temperature ranges for two dislocation relaxation
mechanisms. The structure shifts to the left and vanishes for
higher probe powers, indicating rearrangement of the defect
structure due to stresses or local heating of the nano-specimen.}
\label{figure:4}
\end{figure}

Here, we present experiments where surface acoustic waves (SAW)
are used for the mechanical excitation of nano crystals. By
directly probing the resonance frequency and the quality factor of
a nano resonator, we are able to investigate the influence of SAW
induced strain on the internal friction behavior of single defects
in such a small mechanical system. Following the routes of
classical experiments the strain dependent internal friction
(ADIF) together with its typical temperature dependence indicate
the prominent energy dissipation mechanism being mediated by line
defects. Due to the small size of our nano resonator, we observe a
non-classical step like increase of the damping in the ADIF
experiments, which is attributed to single defects.
%%%%%%%%%%%%%%%%%%%%%%%%%%%%%%%%%%%%%%%%%%%%%%%%%%%%%%%%%%%%%%%%%%%%%
Surface acoustic waves are perfectly suited to provide an
excitation mechanism for probing attenuation effects in mechanical
nano structures. SAW are acoustic modes propagating at the surface
of an elastic solid, having  wavelengths in the range of some
microns and penetration depths of the same order. Excitation of
SAW at a specific frequency is in particular effective employing a
piezoelectric substrate (GaAs in our case) and properly designed
interdigitated transducers (IDTs)~\cite{Datta:1}. Then, a coherent
acoustic ultrasonic beam is generated at the IDT's resonance
frequency $f_{\rm saw} = v_{\rm saw}/\lambda_{\rm saw}$. Here,
$v_{\rm saw} \sim 3$~km~s$^{-1}$ denotes the SAW velocity and
$\lambda_{\rm saw}$ the acoustic wavelength, being
lithographically defined by the spacing between the fingers of the
IDTs (cf. Fig.~\ref{figure:1}). As typical SAW amplitudes can
range in the order of some 10~\AA, they effectively modulate the
acoustic properties of nano mechanical devices~\cite{Beil:1}. The
freely suspended nano crystals are prepared by several successive
steps of electron beam lithography, anisotropic reactive ion
etching (RIE), and selective isotropic wet etching. The details on
production of nano mechanical devices is described in the
literature~\cite{Scheible:1}.
\begin{figure}
\includegraphics[width=8.5cm]{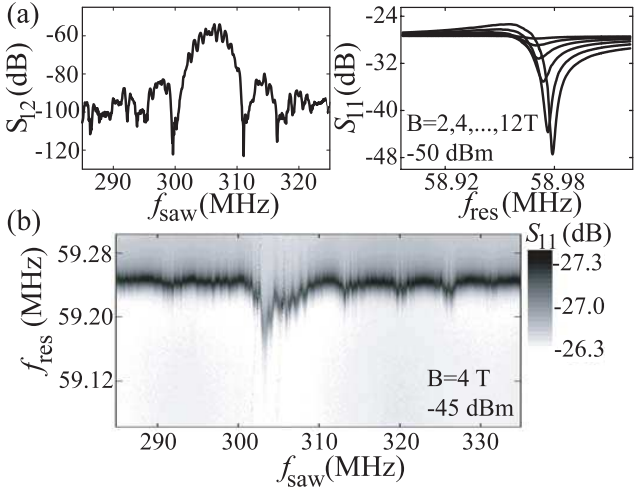}% Here is how to import EPS art
\caption{(a) left: Transmission characteristics of the acoustic
delay line formed by the two IDTs depicted in Fig.~\ref{figure:1}.
The SAW has a maximum amplitude at 305~MHz. right: Magnetomotive
impedance spectroscopy of the resonator mode under investigation
without acoustic loading, as described in the text. Decreasing the
magnetic field decreases the peak in the reflected power, proving
the mechanical origin of the resonance. (b) SAW induced changes in
the magneto impedance signal of the nano beam. Interaction with
the SAW leads to a resonant shift of the eigenfrequency and a
decreasing quality factor of the nano crystal. The SAW power is
25~dBm, while a power of -45~dBm at a magnetic field of 4~T was
applied to the beam. } \label{figure:2}
\end{figure}
%Starting from
%a GaAs/AlGaAs/GaAs heterostructure, in a first step the nano beam,
%the electrical leads, and IDTs are defined and covered by a 80~nm
%thick Au layer. This metal film serves both as a conducting layer
%as well as an etch mask for the subsequent anisotropic RIE step,
%producing the vertical side walls of the nano beam (for details,
%see Ref.~\cite{Scheible:1}). The IDTs and the leads are protected
%by an electron beam resist (PMMA) during this step. In a following
%step, the 50~nm thick AlGaAs sacrificial layer is selectively
%removed by wet etching in 0.1\% hydrofluoric acid (HF), leaving a
%free standing nano resonator.
On the right of Fig.~\ref{figure:1}, we depict a top view of the
setup together with an electron beam micrograph of the sample's
geometry. In our experiments, we first probe the quality factor of
the resonator by standard magnetomotive excitation at the expected
eigenfrequency. Here, an alternating current of the appropriate
frequency is passed through the metallic top layer of the beam.
The sample is placed in a strong dc magnetic field which induces a
periodic Lorentz force acting on the beam. Its response is
detected via standard impedance spectroscopy, and the quality
factor $Q$ of the resonance is determined by the relation $Q =
\omega_{0}/\Delta \omega$ where $\omega_{0}$ is the center
frequency of the resonance and $\Delta \omega$ denotes the full
width at half maximum. For low internal friction this procedure is
justified, as can be shown by fitting to an appropriate equivalent
circuit~\cite{Beil:2}. The interaction between the nano resonator
and the SAW travelling underneath it (cf. Fig.~\ref{figure:1}) is
probed by measuring the changes of the magnetomotively induced
signal. The SAW in this case is excited at a considerably higher
frequency $f_{\rm saw} \sim 300$~MHz than the beam, excluding
cross talk of the IDTs to the resonator setup [cf.
Fig.~\ref{figure:2}(a), left]. The response of the nano beam to
magnetomotive excitation at 59~MHz is shown in
Fig.~\ref{figure:2}(a, right) for different magnetic fields. As an
indication for the interaction between the beam and the SAW, we
observe a considerable shift of the nano mechanical resonance
frequency, a decreased quality factor, and a strong attenuation of
the resonance. This can be seen from Fig.~\ref{figure:2}(b), where
we plot the resonator's driving frequency versus the driving SAW
frequency and code the depth of the resonance in a gray scale
plot. In Fig.~\ref{figure:3}(a) the SAW-attenuation of the
resonance and variation of the eigenfrequency on applied SAW power
are shown. To ensure the SAW induced mechanical origin of the
detuning of the beam resonance, any spurious effects have to be
excluded: Pure heating of the crystal by the SAW can be shown to
be a small effect~\cite{Beil:1, Wixforth:1}. A SAW induced
increase of clamping losses would be accompanied by a a larger
shift in the beam's eigenfrequency, due to a change in the form of
the beam's eigenmode. Induced currents or coupling of
piezoelectric fields can also be discarded, since again they are
generated at $f_{\rm saw}=300$~MHz, whereas the signal is detected
at the beam's eigenfrequency. Hence, the modulation of the
resonator frequency and of the internal dissipation processes are
attributed to a SAW mediated deformation of the oscillating beam.
The responsible SAW induced strain in the nano crystal can be
calculated by solving the free beam equation for the transversal
amplitude of motion, taking into account time dependent boundary
conditions, representing a SAW induced motion of the clamping
points (cf. Fig.~\ref{figure:1}, lower left). For an elliptically
polarized SAW of the Rayleigh type, the clamping points exhibit an
elliptic motion in the sagittal plane, which can be expressed by
the displacement vector
\begin{displaymath}
\boldsymbol{{\rm u}} _ {\rm clamp}(x,t)=
\begin{pmatrix}
A_{\rm l} \\
A_{\rm t} {\rm exp}(i \pi / 2)\\
0
\end{pmatrix}
{\rm exp}(i (k x - \omega_{\rm saw} t)),
\end{displaymath}
where $A_{\rm l}$, and $A_{\rm t}$ denote the longitudinal and
transversal components of the SAW propagating in the $x$ direction
with frequency $\omega_{\rm saw}$ and wave vector component $k$.
The $x$-axis is defined along the [011] direction, whereas the
$y$-axis is perpendicular to the sample surface. Solving the beam
equation, we obtain the following expression for the SAW induced
periodic strain $\epsilon_{ij}$ in the beam extending from 0 to
$L$ along the $x$-axis:
\begin{gather*}
\epsilon_{11}(x)=\frac{A_{\rm l} ({\rm exp}(ikL)-1)}{L}, \\
\epsilon_{12}(x) = \epsilon_{21}(x)=3 {A}_{\rm t} ({\rm
exp}(ikL)-1)(\frac{x}{L^{2}} -\frac{x^{2}}{L^{3}}),
\label{equation 3}
\end{gather*}
where $L$ is the length of the beam. The corresponding SAW
amplitude can be calculated in terms of the SAW power and the IDT
geometry by
\begin{displaymath}
A_{{\rm (l, t)}} = \frac{k e_{ij}}{c_{mn} k^{2} - \rho \omega_{\rm
saw}^{2}} \sqrt{\alpha_{({\rm l,t})} k \frac{P_{\rm saw}}{ 2 \pi
W}},
\end{displaymath}
where $c_{mn}$, and $e_{ij}$ are the appropriate components of the
mechanical and electrical compliance tensor, respectively. $W$ is
the aperture of the IDT, $P_{\rm saw}$ is the injected acoustic
power of the SAW, and $\alpha_{\rm (l,t)}$ are constants
connecting the SAW power with the induced piezoelectric
fields~\cite{Datta:1}. An estimated SAW amplitude of 10~\AA~hence
induces a strain of order $10^{-3}$, resulting in a stress of $12
\times 10^{7}$~N~m$^{-2}$. The shear strain induced by
magnetomotive resonant excitation is on the order of $10^{-5}$ for
an amplitude of 10~\AA, whereas the longitudinal strain in this
case is even smaller. %In conclusion, SAW are capable to exert
%forces on the nano specimen orders of magnitude larger than
%conventional techniques.
%\*********************************************
%\Theorie
The SAW induced damping, as shown in Fig.~\ref{figure:3}(a),
increases for increasing SAW amplitude about an order of
magnitude. In parallel, we observe a shift of the beam's resonance
frequency to lower values. Magnifying one the characteristic
traces of Fig.~\ref{figure:3}(a) reveals an interesting detail:
the dramatic reduction of the resonance quality factor is
accompanied by a step like decrease of the resonance frequency and
quality factor [see Fig.~\ref{figure:3}(b)]. As will be confirmed
by temperature dependent measurements below, we attribute the main
contribution to the SAW-induced damping to the polycrystalline Au
layer on top of the GaAs beam~\cite{Pohl:1}. To explain the
dependence of the damping on the SAW induced strain, we follow a
classical string model, developed by Granato and
L\"ucke~\cite{Nowick:1}. This theory treats dislocations as
vibrating strings, which execute damped oscillations under the
action of a periodically applied force. A straight elastic string
is pinned at both ends, the resonant frequencies of the string are
calculated, and the losses associated are summed to yield a final
expression for the total loss. Dislocations which are pinned to
defects are treated as a series of loops becoming gradually
unpinned, and increasing the length of the strings each time the
net force at a pinning site overcomes a certain binding force. In
this Granato-L\"ucke model the damping  increases as $l^{4}$,
where $l$ is the dislocation loop length. For a random
distribution of the pinning sites along the dislocation, Granato
and L\"ucke obtain the following expression for the dependence of
internal friction on tensile strain $\epsilon$:
\begin{displaymath}
Q^{-1}(\epsilon) = (C_{1} / \epsilon) {\rm exp}(-C_{2} / \epsilon)
\tag{1}.
\end{displaymath}
Here, $C_{2} \sim l_{0}^{-1}$ and $C_{1} \sim (\Lambda L_{\rm
dis}^{3} /l_0)C_{2}$. $\Lambda$ is the dislocation density,
$l_{0}$ the average loop length, and $L_{\rm dis}$ is the length
of the dislocation. The theory also predicts an amplitude
dependent modulus defect of the same form as Eq.~(1). This
theoretical description can be tested against experiments when
plotting log($Q^{-1} \epsilon$) versus the reciprocal strain
$\epsilon^{-1}$ and looking for a linear dependence. The slope
then varies inversely with the average loop length while the
intersection with $\epsilon^{-1}=0$ is sensitive to $\Lambda
L_{\rm dis}^{3}/l_0$. These so called 'Granato-L\"ucke plots' are
shown in Fig.\ref{figure:4}(a) for two different ac driving powers
at the beam. As expected, we obtain a linear behavior, although
the occurring step-like modulations indicate a significant
deviation from the classical model. To clarify this point, we note
that the above string model assumes a statistical distribution of
pinning centers and loop lengths. This can only be true for a
large number of dislocations, which is certainly not the case for
our nano beam. Moreover, the elastic energy stored in the
oscillating nano crystal at the highest amplitude is of the order
of 2 $\times 10^{-19}$~J $\sim$ 1~eV. This implies, that single
defects with activation energies as low as 0.1~eV should be
observable in the $Q$-value of the nano crystal resonance. In
these terms, we attribute the steps in the ADIF measurements
[Fig.~\ref{figure:3}(b)] to unpinning events for single
dislocations in the gold layer having different activation
energies. These steps in the Granato-L\"ucke plots tend to
disappear for samples that have considerable more defects and
pinning centers. Figure~\ref{figure:4} additionally shows data
obtained for another sample at 17~MHz with same dimensions, but
increased thickness of the gold layer. The linear quality factor
of this second sample $Q_2=1620$ was considerable smaller than the
equivalent quantity of the first sample $Q_1=2923$. This fact,
together with the slope in the Granato-L\"ucke plot, shows that in
the second sample more dislocations, with smaller average loop
length, i.e. considerable more pinning centers, are present. The
Granato-L\"ucke plots for this measurements don't show a step like
structure, due to a better statistical distribution of point
defects along the dislocations.
In addition to ADIF, dislocation mediated damping exhibits a
characteristic temperature dependence. For macroscopic samples of
fcc metals, typical traces exhibit a broad damping maximum around
120~K, known as the 'Bordoni peak', followed by an increase in
damping~\cite{Marchesoni:1}. In Fig.~\ref{figure:4}(b), we depict
such temperature dependent measurements of the internal friction
for another, similar nano beam oscillating at 61~MHz.
As in this case the eigenfrequency is four orders of magnitude
larger than in macroscopic samples, whereas the maximum
dislocation length is two orders of magnitude smaller the Bordoni
peak is expected to be shifted in temperature, due to the
Arrhenius equation for thermally activated relaxation
processes~\cite{Nowick:1}. Many theories regarding kink-pair
formation assume a quadratic dependence of relaxation time on
dislocation length (see e.g.~\cite{Nowick:1, Brailsford:1}), so
that we expect a relaxation time four orders of magnitude smaller
than in macroscopic samples (around 10$^{-16}$ s). Taking the
known activation energies for the Bordoni peak (around 0.1~eV) and
the Niblett-Wilks peak (around 0.05~eV), we expect damping peaks
due to dislocation relaxation around 30~K and 50~K.
%
%Although in nano crystalline materials a Bordoni peak is observed,
%the mechanisms responsible for its occurrence are thought to be
%related to dislocation structures at grain
%boundaries~\cite{Tang:1}.
Experiments confirm a Bordoni like relaxation peak down to grain
sizes below 30~nm~\cite{Tang:1}. Using an atomic force microscope,
we determined a typical grain size of about 50~nm in the
polycrystalline gold layer for our samples, which is still large
enough for kink-pair formation.
%
%From this, an estimated number of $\sim$ 100 grains forms the
%conducting layer.
%\*********************************************

In summary, we have demonstrated the dynamical tuning of internal
friction in nano crystals by surface acoustic waves. The
evaluation of the SAW induced dissipation in nano beams shows that
dislocations are responsible for the induced acousto-friction. Due
to the  small size of our specimen we are able to resolve the
contribution of single defects to the mechanical loss. Temperature
dependent measurements exhibit a damping peak at T=30~K followed
by an increase in damping. This is explained by kink-pair
formation whereas relaxation at the grain boundaries might play an
additional role~\cite{Tang:1}.

We gratefully acknowledge financial support by the Deutsche
Forschungsgemeinschaft under contract number DFG/Bl-487/3. We also
like to thank J\"org P. Kotthaus for stimulating discussions and
continuing support.

%%%%%%%%%%%%%%%%%%%
%%%%%%%%%%%%%%%%%%%%%%%%%%%%%%%%%%%%%%%%%%%%%%%%%%%%%

\newpage %Just because of unusual number of tables stacked at end
%\bibliography{PRL_SAW_DAMP}% Produces the bibliography via BibTeX.
%%%%%%%%%%%%%%%%%%%%%%%%%%%%%%
%%%%%%%%%%%%%%%%%%%%%%%%%%%%%%
%%%%%%%%%%%%%%%%%%%%%%%%%%%%%%
%%%%%%%%%%%%%%%%%%%%%%%%%%%%%%
%%%%%%%%%%%%%%%%%%%%%%%%%%%%%%
\end{document}